\documentclass[manuscript]{aastex}
\usepackage{graphicx}
\usepackage{url}
\usepackage{color}

\shorttitle{Low-energy $\alpha$-particles in flares}
\shortauthors{Hudson et al.}

\begin{document}

\title{Charge-exchange limits on low-energy \\$\alpha$-particle fluxes in solar flares}

\author{H.~S. Hudson$^{1,2}$, L. Fletcher$^2$, A.~L. MacKinnon$^2$, and T.~N. Woods$^3$}
\affil{$^1$SSL, UC Berkeley, CA, USA 94720; 
$^2$School of Physics and Astronomy, SUPA, University of Glasgow, Glasgow G12 8QQ, UK; 
$^3$Laboratory for Atmospheric and Space Physics, University of Colorado, 1234 Innovation Dr., Boulder,
CO 80303, USA
}
\email{hhudson@ssl.berkeley.edu}


\begin{abstract}
This paper reports on a search for flare emission via charge-exchange radiation in the wings of the Lyman-$\alpha$ line of He~{\sc ii} at 304~\AA, as originally suggested for hydrogen by \cite{1976ApJ...208..618O}.
Via this mechanism a primary $\alpha$ particle that penetrates into the neutral chromosphere can pick up an atomic electron and  emit in the He~{\sc ii} bound-bound spectrum before it stops.
The Extreme-ultraviolet Variability Experiment (EVE) on board the \textit{Solar Dynamics Observatory (SDO)} gives us our first chance to search for this effect systematically.
The Orrall-Zirker mechanism has  great importance for flare physics because of the essential roles that particle acceleration plays;  this mechanism is one of the few proposed that would allow remote sensing of primary accelerated particles below a few MeV~nucleon$^{-1}$.
We study ten events in total, including the $\gamma$-ray events SOL2010-06-12 (M2.0) and SOL2011-02-24 (M3.5) (the latter a limb flare), seven X-class flares, and one prominent M-class event  that produced solar energetic particles (SEPs).
The absence of charge-exchange line wings may point to a need for more complete theoretical work.
Some of the events do have broad-band signatures, which could correspond to continua from other origins, but these do not have the spectral signatures expected from the Orrall-Zirker mechanism.

\end{abstract}


\section{Introduction}\label{sec:intro}

The behavior of high-energy particles in solar flares lies at the core of the physics.
We detect such particles in flares via remote sensing of their bremsstrahlung continuum (for electrons) and via $\gamma$-ray line and continuum emissions (for energetic ions).
The ion-related continua include those from pion decay and the ``nuclear continuum'' composed of unresolved lines.
 Flare energy released suddenly from the coronal magnetic field tends to appear  initially in the form of high-energy particles \citep[e.g.,][]{1976SoPh...50..153L,1995ApJ...455L.193R}.
Unfortunately, though, the X-ray and $\gamma$-ray emissions are not generally sensitive to non-thermal electrons at energies below about 10~keV, or protons below about 20~MeV, respectively.

\cite{1976ApJ...208..618O} proposed one of the few plausible techniques capable of filling in this major observational gap, namely the charge-exchange line radiation that could be created by lower-energy particles.
 Historically, we note that the proton aurora was first identified via a similar technique: the observation of Doppler-shifted H$\alpha$ radiation  \citep{1939Natur.144.1089V,1967RvGSP...5..207E}.
In this mechanism a fast proton penetrates a region of low ionization, picks up an electron, and then radiates a Doppler-shifted emission-line spectrum.
For a solar proton moving directly away from the observer, i.e., towards the photosphere from a coronal source at disk center (assuming {\bf B} radial), the hydrogen Lyman-$\alpha$ line would appear at 
 5-30~\AA~to the red, for proton energies of 10-100~keV \citep{1976ApJ...208..618O}.
A more complicated beam  or field geometry would generally produce both red and blue line wings and alter the spectral mapping, but in the idealized process the red wing dominates.
Charge-exchange line wings will also develop for other lines, such as the He~{\sc ii} Lyman-$\alpha$ line at 304~\AA~that we study in this paper. 

The intensity of the Lyman-$\alpha$ charge-exchange radiation depends upon several factors, as discussed further in several papers \citep{1985ApJ...295..275C,1995ApJ...441..385B,1985ApJ...289..425F,1999ApJ...514..430B}.
First one must assume the existence of a proton beam in the first place; the particles producing the $\gamma$-ray emission may have broad pitch-angle distributions \citep[see][for a review of proton effects in solar flares]{1995SSRv...73..387S}.
The intensity of the beam obviously dictates the intensity of the emission; \cite{1976ApJ...208..618O} assumed a proton beam flux of only $4 \times 10^7$~erg~cm$^{-2}$~s$^{-1}$, a weak flare, whereas \cite{1985ApJ...295..275C} extended the theory to a proton beam intensity comparable in energy to that assumed to be present in a major flare ($\sim$$10^{11}$~erg~cm$^{-2}$~s$^{-1}$).
The protons may induce wave-particle interactions \citep{1989ApJ...342..576T} that can compete with the more standard theory \citep[e.g.,][]{1985ApJ...295..648E}, based on Coulomb collisions, that is often applied to problems of particle propagation in solar flares.
Alternatively they may be a part of a neutral beam \citep[e.g.,][]{1990ApJS...73..333M}.
For the charge-exchange mechanism to work, the high-energy ions must propagate in a sufficiently un-ionized medium, and yet not ionize it, so that the charge-exchange process remains efficient; the ionization of the plasma requires a finite time interval that may compete with the time scales for flare development.
Detailed calculations such as those of \cite{1985ApJ...295..275C} invoke equilibrium conditions between the beam and the target, but also discuss the time scales over which heating (which greatly reduces the intensity of the charge-exchange emission) results from the beam interactions themselves.

 The helium line wings produced by charge-exchange reactions were mentioned by \cite{1976ApJ...208..618O}, and then calculated in detail by \cite{1990ApJ...351..317P}.
The ions accelerated in a solar flare almost certainly include a proportion of $\alpha$~particles \citep{1997ApJ...485..409S}; the He~{\sc ii} wing intensity will depend the $\alpha$/proton ratio, which \cite{1998ApJ...508..876S} find to be significantly larger than for the ambient photosphere, perhaps more than 50\%.
As with the Ly$\alpha$ charge-exchange radiation, the signal should appear as broad wings of the line, with the peak red (or blue) wavelength determined by the assumed decrease of primary particle flux (at large $\Delta\lambda$), and the diminishing primary particle range against collisional losses (at small $\Delta\lambda$).

The EVE instrument on \textit{SDO} opens a new window of opportunity to detect and exploit the charge-exchange line wings, owing to its temporal and spectral resolution.
It provides 1~\AA~spectral resolution, at 10~s cadence, for the total solar spectrum \citep{2012SoPh..275..115W}.
Its MEGS-A (Multiple EUV Grating Spectrograph-A) spectrometer, which we use here, covers the range 64--370~\AA.
Some earlier searches for the charge-exchange mechanism have been possible, but not with comprehensive coverage nor for major events known to have $\gamma$-ray emission.
\cite{2001ApJ...555..435B} studied a GOES C3.8 event with CDS data, obtaining no detection but setting a limit of 250 erg~cm$^{-2}$~s$^{-1}$~sr$^{-1}$~\AA$^{-1}$.
EVE is a Sun-as-a-star instrument, and thus suffers from relatively high background levels due to non-flare emission from the rest of the Sun.
Nevertheless solar flares produce clear signatures  across the spectrum \citep{2010AGUFMSH13A..01W}, with good signal-to-noise ratio for the He~{\sc ii} 304~\AA~line in major M- or X-class flares \citep[e.g.,][]{2011SoPh..273...69H}.
This latter paper also describes the Doppler sensitivity of the MEGS-A spectra.

In this paper we describe searches for broad wings of the 304~\AA~line for a set of ten flares, including seven X-class events, plus two others with $\gamma$-ray emission detected by \textit{RHESSI}\footnote{The \textit{Reuven Ramaty High Energy Spectroscopic Imager} spacecraft; \citep[see][]{2002SoPh..210....3L}}
and \textit{Fermi} \citep{2010AAS...21640406S}.
We describe the EVE data in Section~\ref{sec:data} and the analysis in Section~\ref{sec:deeper}, and then discuss what the observations imply -- upper limits on the charge-exchange wings, but other interesting features -- in Section~\ref{sec:discussion}.

\section{Data survey}\label{sec:data}

\begin{figure}[h]
\plotone{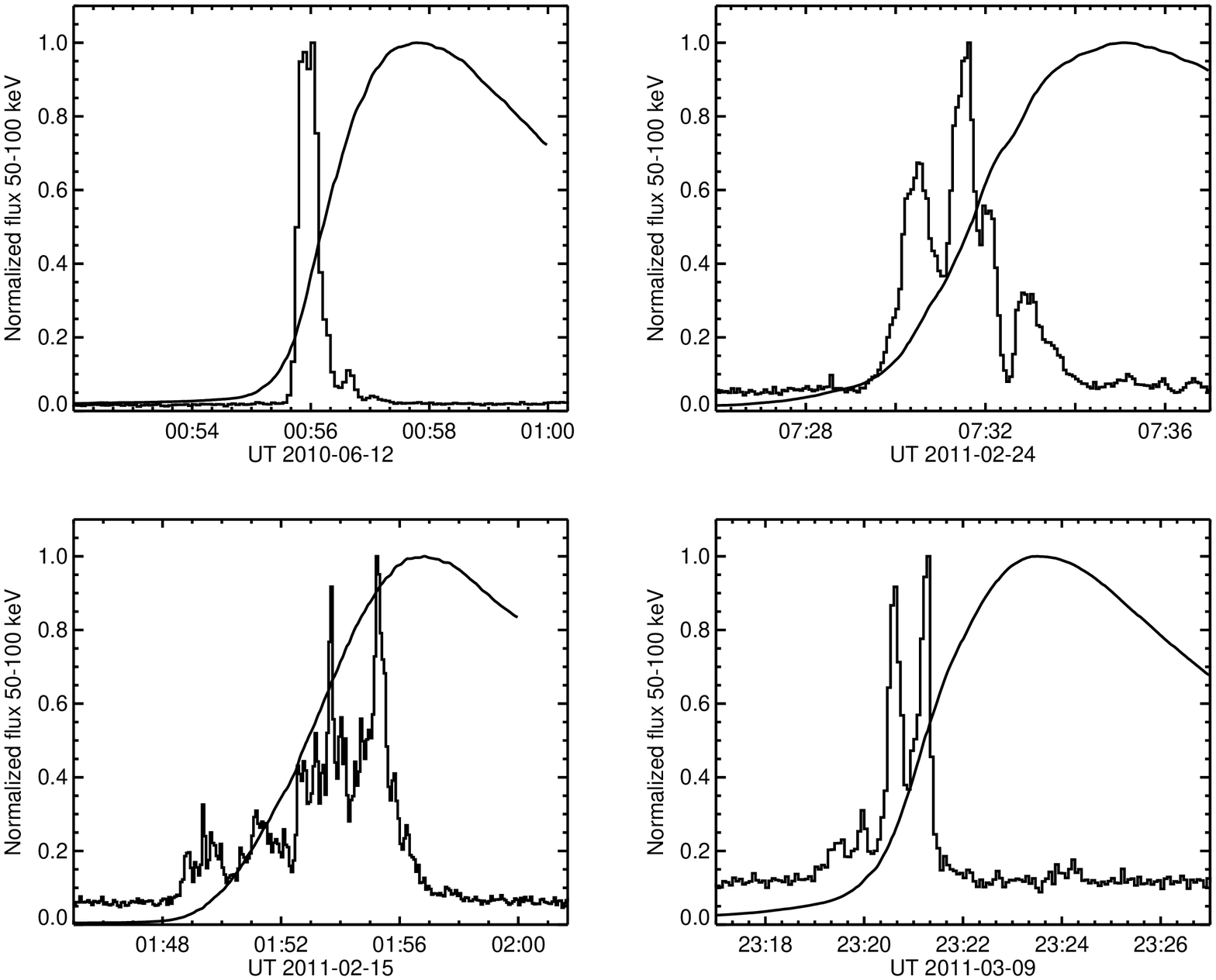}
\caption{The \textit{RHESSI} 50-100~keV and \textit{GOES} 1-8~\AA~time profiles for the four primary flares, each trace
normalized to unity.
The upper row shows the $\gamma$-ray flares, and the lower row the X-class flares.
 In each case the jagged trace shows the hard X-rays and the smooth one, the soft X-rays.
}\label{fig:4x}
\end{figure}

The flares studied, as listed in Table~\ref{tab:flares}, include all of the $\gamma$-ray events and X-class flares
observed after EVE launch 2010 February~11, up to the time of writing.
The first four events in the table are presented in some of the four-panel figures below, and the whole set in ten-panel plots.
The Table shows that EVE successfully observed all of the events, and that both of the  $\gamma$-ray observatories (\textit{RHESSI} and \textit{Fermi}) observed seven of them.
Only SOL2011-09-22T11:01 was not observed at high energies, and of the others eight (those marked with ``$\gamma$'' in the table) had some signature of the presence of hard X-rays above 100~keV, or $\gamma$-ray photons, and hence by implication involved particle acceleration to high energies.
The presence of accelerated $\alpha$~particles in these events is hypothetical, but highly likely based on earlier experience \citep{1997ApJ...485..409S}.
The M-class event SOL2011-08-04T03:57 (M9.7) was notable because it produced the strongest\footnote{http://umbra.gsfc.nasa.gov/SEP/}
SEP event of Cycle 24  at the time of writing, though at an intensity ($>$10 MeV) of only about 100~particles~(cm$^2$ s sr)$^{-1}$ much smaller than many SEP events of the previous cycle.
Finally, two of the events (SOL2011-02-24T07:35 and SOL2011-09-22T11:01) were limb events, with the latter being the weakest HXR/$\gamma$-ray event in the nine-event sample.
Figure~\ref{fig:4x} shows the \textit{RHESSI} hard X-ray and \textit{GOES} soft X-ray time profiles for each of the four primary events.

\begin{table*}[htpb]
\caption{Flares Studied}
\smallskip
\begin{tabular}{| l | l | l | l | l | }
\hline
Flare Identifier & Location & \multicolumn{3}{c|}{Observations} \\ \cline{3-5}
 & & EVE & \textit{RHESSI} & \textit{Fermi} GBM \\
\hline
\hline
SOL2010-06-12T00:57 (M2.0) & N23W43 &Y & Y, $\gamma$ & Y, $\gamma$\\
SOL2011-02-15T01:56 (X2.2) & S20W15 &Y & Y, $\gamma$ &  Y \\
SOL2011-02-24T07:35 (M3.5) & N17E87& Y& Y, $\gamma$ &Y, $\gamma$\\
SOL2011-03-09T23:23 (X1.5) & N08W09 &Y & Y & Y \\
\hline
SOL2011-08-04T03:57 (M9.3) & N19W36 & Y& N & Y, $\gamma$ \\
SOL2011-08-09T08:05 (X6.9) & N17W69 &Y &Y, $\gamma$ & Y, $\gamma$ \\
SOL2011-09-06T22:20 (X2.1) &N14W18 & Y &Y, $\gamma$ & Y, $\gamma$\\
SOL2011-09-07T22:38 (X1.8) & N14W28 & Y&N& Y, $\gamma$\\
SOL2011-09-22T11:01 (X1.4) & N15E83 & Y&N&N\\
SOL2011-09-24T09:40 (X1.9) & N12E60 &Y&Y, $\gamma$&Y, $\gamma$\\
\hline
\end{tabular}\label{tab:flares}
\end{table*}

EVE obtains spectra with 10-s time integrations and approximately 1~\AA~spectral resolution in two units.
The  Multiple EUV Grating Spectrograph-A (MEGS-A) grazing-incidence spectrograph covers the 50-370~\AA~range, with the spectrum presented on a CCD detector \citep{2010SoPh..tmp....3W}.
The resulting signal-to-noise ratio is extremely good, and this gives great freedom for analysis of the charge-exchange line wings, with sensitivity limited only by the photon statistics of the background counting rates.
Emission from the non-flaring Sun dominates these background rates, which vary slowly  in the absence of
competing flares.
Figure~\ref{fig:heii} illustrates the time behavior of the EVE data for SOL2011-02-15T01:56, the most powerful event
of the four events chosen for primary analysis (see Figure~{\ref{fig:4x}, lower left panel, for comparison with the \textit{RHESSI} and \textit{GOES} time series).
The He~{\sc ii} line has a strong impulsive-phase component, but not in excess of the gradual component as in the
case of SOL2010-06-12 \citep{2011SoPh..273...69H}.
 The Figure shows the time variation of spectral irradiance at the nominal wavelength, and the spectrum around that wavelength, without (upper panels) and with (lower panels) subtraction of a pre-flare sample.
The nominal centroid wavelength of the doublet is taken from the CHIANTI database \citep{1997A&AS..125..149D,2009A&A...498..915D}.

\begin{figure}[htpb]
\plotone{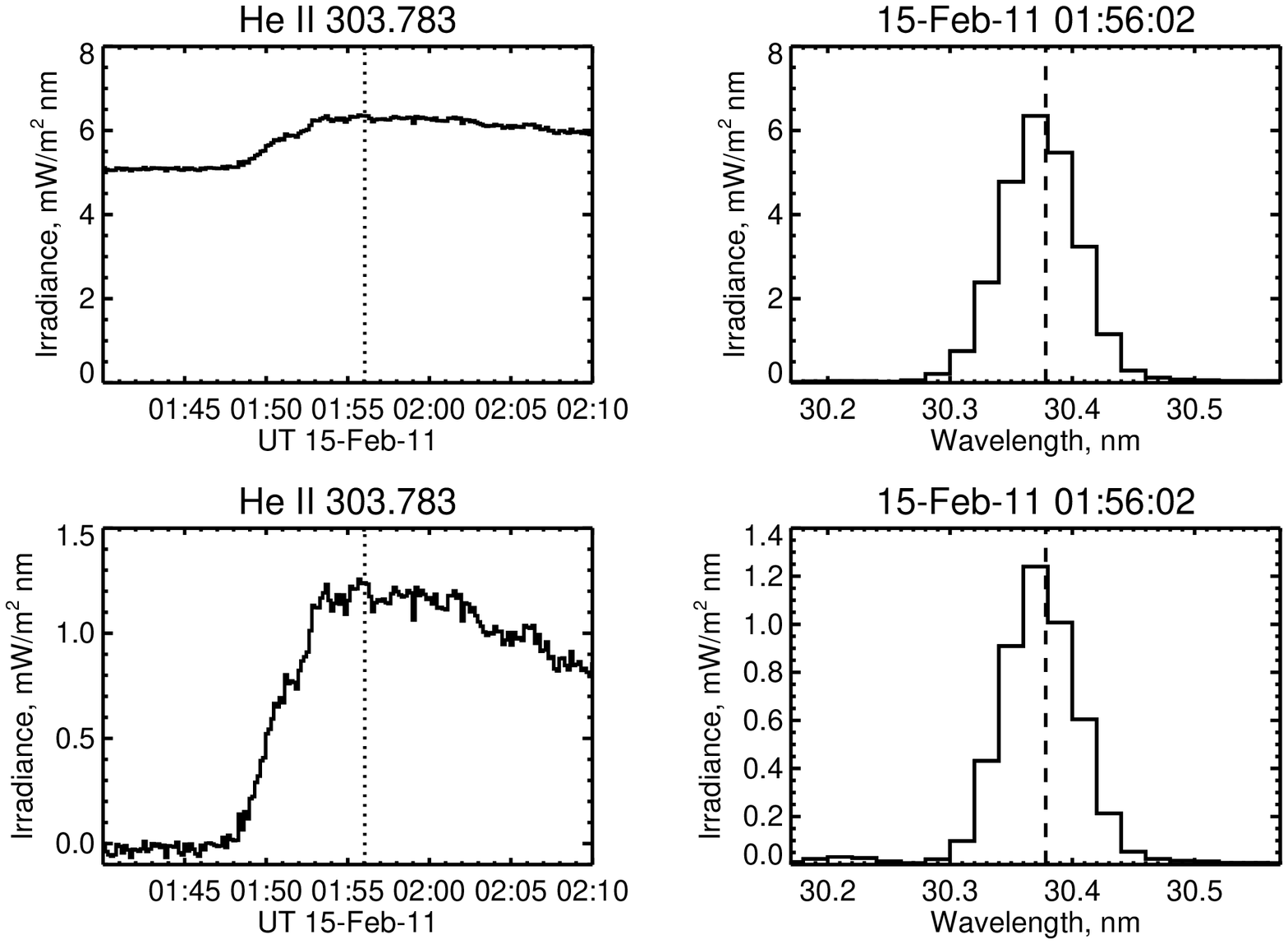}
\caption{Overview of the behavior of the He~{\sc ii}~30.4~nm~line during the flare SOL2011-02-15T01:56 (X2.2). 
Left: time series of the excess emission at line peak relative to a preflare interval  (see Table~\ref{tab:times}).
Right: the line profile at GOES maximum.
The flare resulted in an increase of about 20\% in the total 30.4~nm~flux in total.
The dotted line  (left) shows the time of \textit{GOES} maximum, showing that the 30.4~nm emission appeared in the impulsive phase as well as later on.
 For the lower plots a (single) background integration at 01:40~UT was subtracted both from the time series (left, for a single MEGS-A spectral bin at the line peak wavelength), and the spectrum (right), shown for the GOES peak time of 01:56~UT.
The dashed line  (right)  shows the nominal mean wavelength of the doublet as listed in CHIANTI.
}\label{fig:heii}
\end{figure}

Figure~\ref{fig:heii} and especially Figure~\ref{fig:4backgrounds} show how stable the EVE/MEGS-A spectra can be.
The latter plots the background spectra used for  all of the events studied. 
The small differences in level result from global changes of the spectral irradiance in the vicinity of the line, given that the flares are spread out over many months.
The background for SOL2010-06-12 is lower than that for the 2011 events, consistent with the solar-cycle increase.
The MEGS-A stability enables good use to be made of the the excellent counting statistics in the 10-s integrations, for example in constructing difference spectra with good precision as shown in Figure~\ref{fig:heii}, lower-right panel.

\begin{figure}[htpb]
\plotone{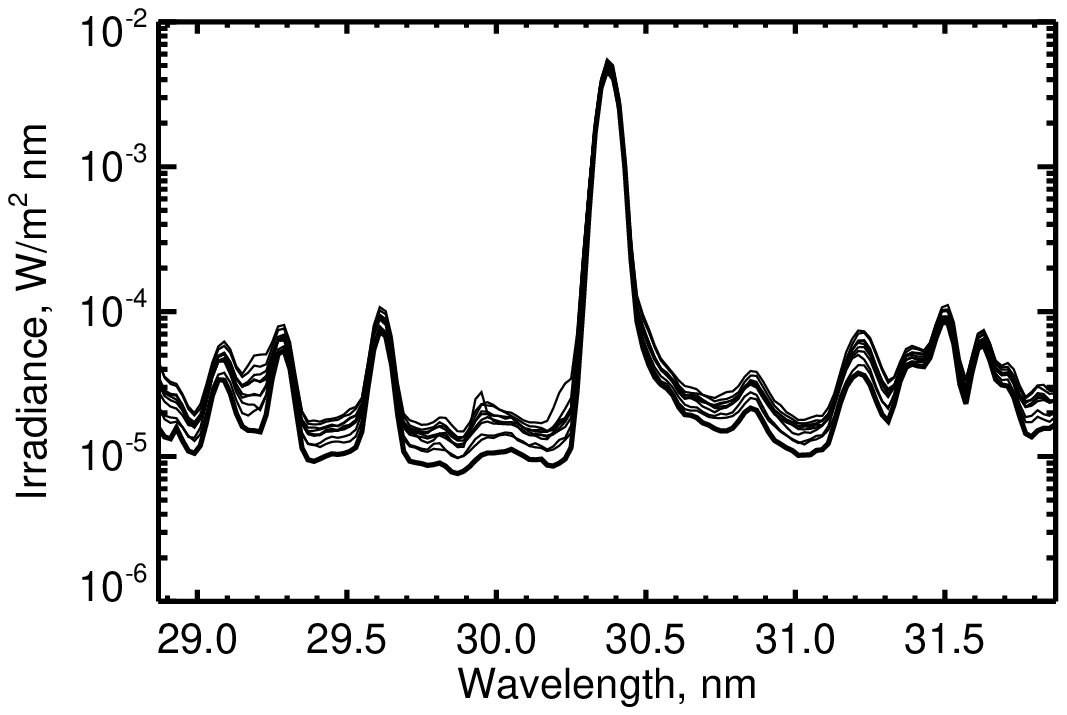}
\caption{EVE MEGS-A background spectra for the ten flares.
SOL2010-06-12 is the heavy line (the lowest).
Table~\ref{tab:times} lists the integration times, each for a 110-s interval,  for these spectra.
See the online version of the paper for color.
}
\label{fig:4backgrounds}
\end{figure}

\begin{table*}
\caption{Time intervals}
\smallskip
\begin{tabular}{| r | r | r | r | }
\hline
Background & Start & Peak & End \\
 \hline
12-Jun-10 00:52& 12-Jun-10 00:55& 12-Jun-10 00:57& 12-Jun-10 01:07\\
15-Feb-11 01:45& 15-Feb-11 01:48& 15-Feb-11 01:57& 15-Feb-11 02:20\\
24-Feb-11 07:25& 24-Feb-11 07:30& 24-Feb-11 07:35& 24-Feb-11 07:50\\
09-Mar-11 23:17& 09-Mar-11 23:19& 09-Mar-11 23:23& 09-Mar-11 23:29\\
04-Aug-11 03:35& 04-Aug-11 03:44& 04-Aug-11 03:57& 04-Aug-11 04:04\\
09-Aug-11 07:55& 09-Aug-11 08:00& 09-Aug-11 08:08& 09-Aug-11 08:15\\
06-Sep-11 22:12& 06-Sep-11 22:16& 06-Sep-11 22:25& 06-Sep-11 22:40\\
07-Sep-11 22:30& 07-Sep-11 22:34& 07-Sep-11 22:40& 07-Sep-11 22:48\\
22-Sep-11 10:45& 22-Sep-11 10:50& 22-Sep-11 11:01& 22-Sep-11 11:11\\
24-Sep-11 09:32& 24-Sep-11 09:34& 24-Sep-11 09:40& 24-Sep-11 09:48\\
\hline
\end{tabular}\label{tab:times}
\end{table*}

Figures~\ref{fig:4panel_cont} and~\ref{fig:4panel_ts} show a search for the beam signature in the four primary events.
According to the \cite{1990ApJ...351..317P} calculations, a broad red wing should appear, with a peak wavelength depending on the $\alpha$-particle energy.
 The results of \cite{1997ApJ...485..409S} for other $\gamma$-ray events suggest that the particle beams may be diffusive, rather than highly directional. 
Thus the blue wing may also be enhanced as a result of upward particle motions. 
 Note that such motions
are not a part of the simple Orrall-Zirker theory, but would be expected in a realistic field structure due to mirroring
and tilts, as well as to pitch-angle scattering; see e.g. \cite{2007ApJS..168..167M} or \cite{2008A&A...489L..57K}.
This search therefore defined three bands (one blue, two red) in the line wings, as shown by the dotted lines in 
Figure~\ref{fig:4panel_cont}. 
These locations correspond to weak line  emission, as determined by the appearance of the spectra as 
well as the predictions from the CHIANTI models (see below).
The widths of the chosen regions correspond to the MEGS-A spectral resolution \citep{2010SoPh..tmp....3W}.
The time series of excess irradiance in these bands, for each flare, then appears in Figure~\ref{fig:4panel_ts}.
A wavelength of 31~nm (as shown) corresponds to 0.5~MeV/amu in the initial beam spectrum; at higher energies the charge-exchange cross-section for $\alpha$~particles on H~atoms rapidly decreases.
No signal with the expected distribution is present; the models predict enhanced wing(s) with peaks separated from
the line core by an amount larger than the MEGS-A resolution \citep{1990ApJ...351..317P}.
We note the appearance of a weak flare-associated excess in the search area of the spectrum,  in both the impulsive and gradual-phase spectra shown, but that it does not have the spectral profile expected for the broad line wings.
For example, there is a comparable blue-wing excess for SOL2011-02-15, which would imply upward particle motions
even for an event near disk center (S20W15).

Any of the excess fluxes could well be due to unresolved spectral lines or other continuum sources \citep[e.g.,][]{1987aeus.book.....F}.
 The evaporation blueshifts known to occur in the impulsive phase of a flare, as described for EVE data by \cite{2011SoPh..273...69H}, do not extend as far as our blue reference band, and would thus not be detectable.
We discuss other possibilities in Section~\ref{sec:discussion}.

\begin{figure}[htpb]
\plotone{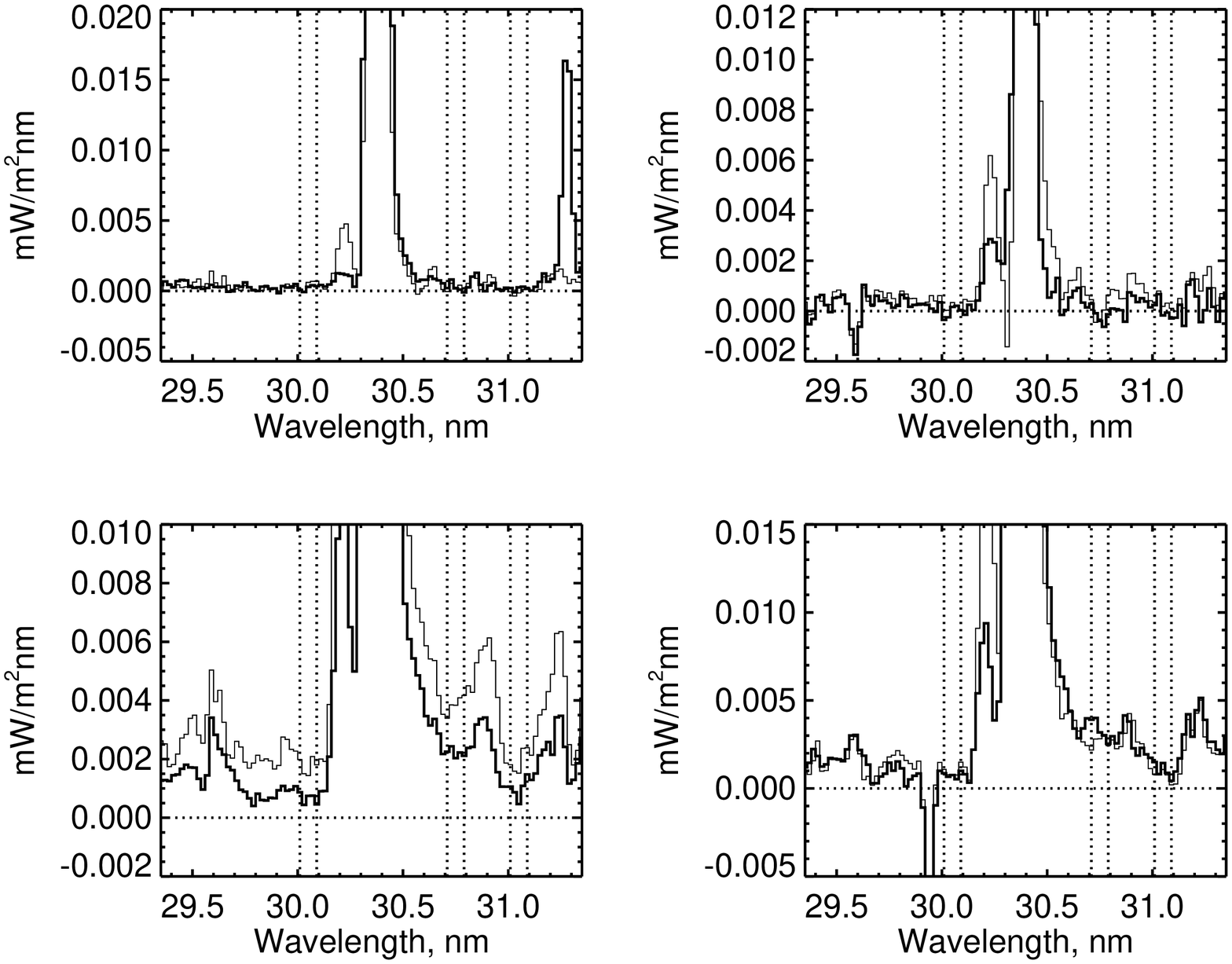}
\caption{The EUV spectra for the four primary flares, in the order shown in Figure~\ref{fig:4x}.
In each case the heavy line shows the excess spectrum during the impulsive phase, and the light line the gradual phase, as defined in Table~\ref{tab:times}  (which also gives the times for the background integrations).
From left to right, the dotted lines show the blue-1, red-1, and red-2 bands.
}\label{fig:4panel_cont}
\end{figure}

\begin{figure}[h]
\plotone{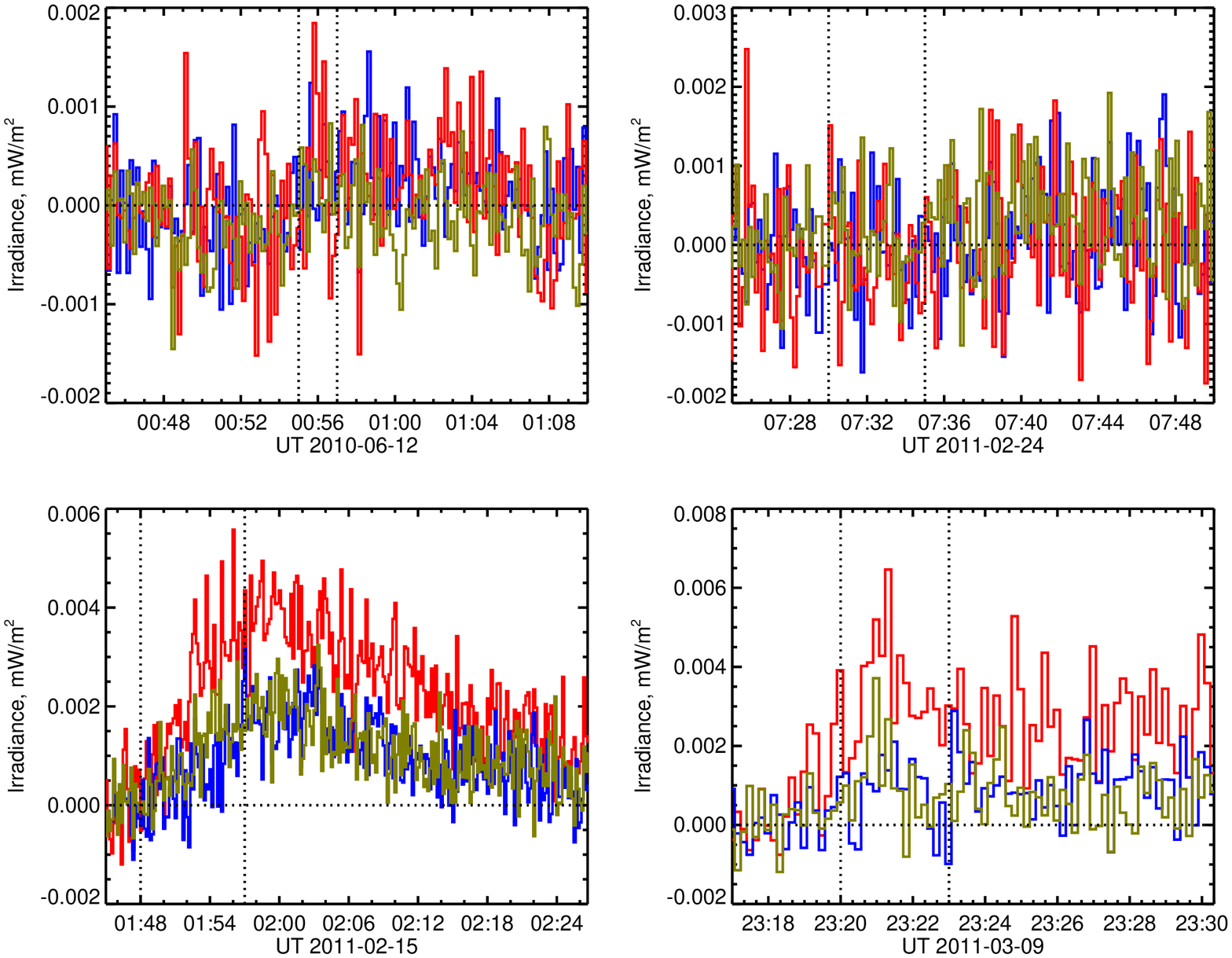}
\caption{The EUV time series for each spectral band, in the order shown in Figure~\ref{fig:4x}.
Red-1 is in red, red-2 in gold, and blue-1 in blue.
The vertical dotted lines show the time ranges for each impulsive phase.
}\label{fig:4panel_ts}
\end{figure}

The observed spectra make it clear that MEGS-A observes many lines, and that they tend to dominate the
continuum.
We have used CHIANTI \citep{1997A&AS..125..149D,2009A&A...498..915D} to get a feeling for the line population in the domain of the He~{\sc ii} charge-exchange regions. 
The spectrum in Figure~\ref{fig:chianti} shows a CHIANTI model with standard assumptions of coronal abundances and the \cite{1998A&AS..133..403M} ionization equilibrium, smoothed to the spectral resolution of MEGS-A, in comparison with
the integrated excess spectra observed for SOL2010-06-12.
This model uses the flare emission-measure distribution (DEM) of \cite{1979ApJ...229..772D}, as extended to high temperatures by G. Del Zanna.

\begin{figure}[h]
\plotone{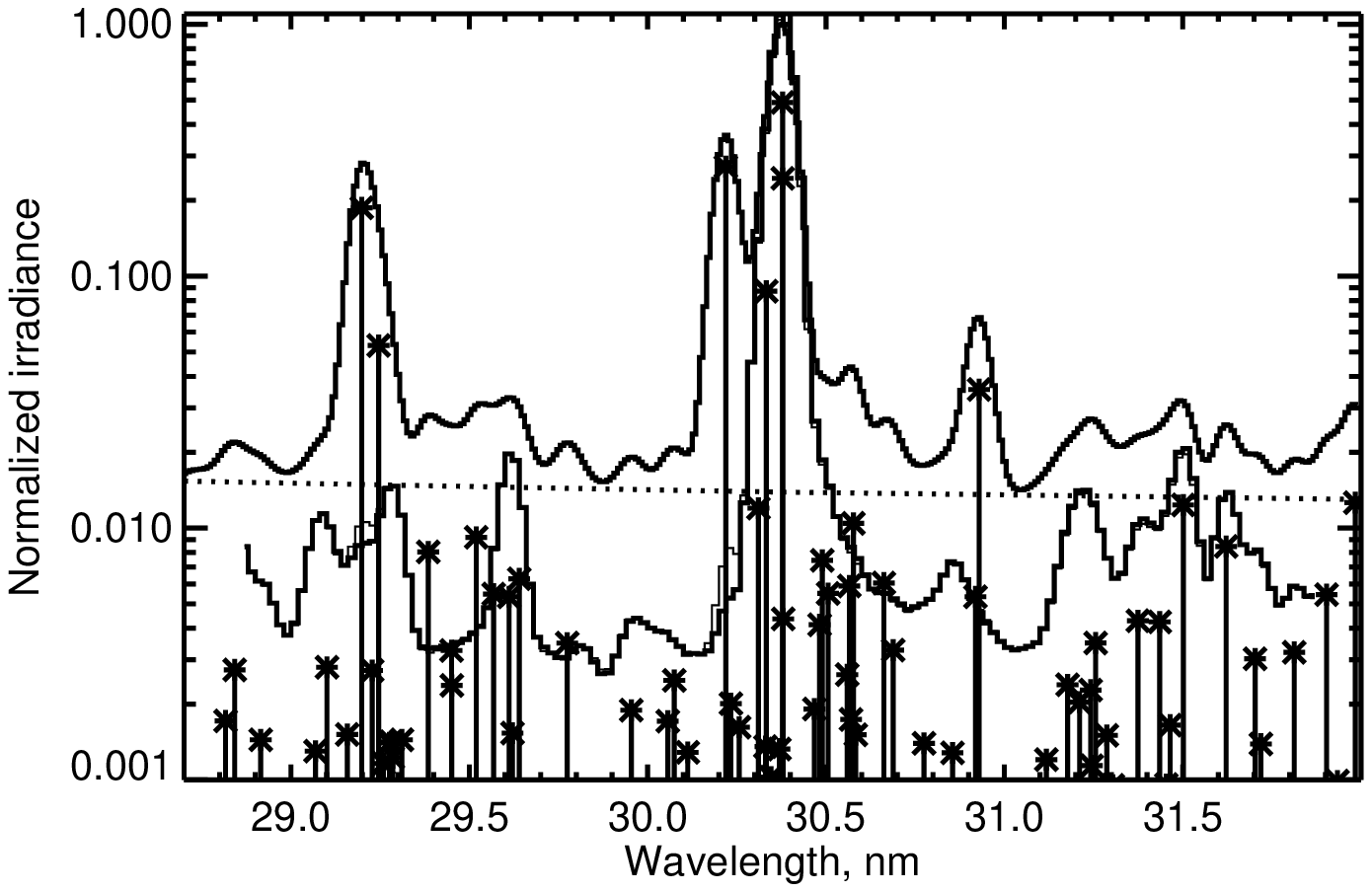}
\caption{CHIANTI model spectrum smoothed to MEGS-A resolution (black histogram), with the strong lines (asterisks) shown for the region of the charge-exchange  line wings; the dotted line shows the CHIANTI model continuum, based on the flare DEM of \cite{1979ApJ...229..772D}.
The model uses the coronal abundances of \cite{1992ApJS...81..387F} and the \cite{1998A&AS..133..403M} ionization equilibrium.
The lower histograms are the impulsive-phase (heavy) and gradual-phase (light, hard to see) integrations  (excess relative to
the selected background interval) for SOL2010-06-12.
Note that the line populations show both agreements and differences  (e.g., CHIANTI predicts a Ca~{\sc xviii} line at 30.22~nm, which is not strong in either phase of this flare).
See the online paper for a color version of this figure.
}\label{fig:chianti}
\end{figure}

In studying Figure~\ref{fig:chianti}, we note that the normalization to the peak 30.4~nm irradiance predicts a spectrum that is low compared with other components of the observed spectrum.
We can also see discrepancies in the line population: Ca~{\sc xviii} (30.219~nm) is overstated in the model, and
we find this to be true of all ten events in our survey.
This is beyond the scope of the present paper, but clearly the EVE spectra will produce more definitive emission-measure distributions for major flares than CHIANTI's representative flare DEM.
As regards the continuum, the CHIANTI model points to locations in the spectrum where it may dominate.
 Note that the impulsive-phase and gradual-phase spectra do not differ by very much in this representation, i.e.,
the red and blue lines almost overlap.
This overlap conceals the fact that the excellent signal-to-noise ratio of EVE allows them to be distinguished quantitatively \citep[e.g.,][]{2011SoPh..273...69H}.
We return to this in Section~\ref{sec:discussion}.

\section{Analysis of ten events}\label{sec:deeper}

These initial four events show no obvious sign of the charge-exchange line wings, but data from the other major events
are available.
We therefore display all of this information in the ten-panel plots of Figures~\ref{fig:ten_imp} and~\ref{fig:ten_grad},
which represent integrations over the impulsive and gradual phases of each event.
We have defined these time intervals by referring to the GOES event times, with adjustments to match the 30.4~nm
event time histories as seen by EVE/MEGS-A, which differ in some details.
Table~\ref{tab:times} lists the times used for the integrations.
In general, a simple background subtraction from a pre-flare level should suffice to show the newly-appearing emission features associated with a flare.
In practice, though, there may be systematic uncertainties regarding this.
As \cite{1990ApJ...356..733B} pointed out in the context of whole-Sun soft X-ray observations, if a flaring structure had previously been contributing to the background level and were inadvertently subtracted, this would result in an overestimate of the flare brightness in that feature.
Also, EUV radiation generally can easily be absorbed by cool material such as that often found near active regions
in prominences.
In these ten cases, there are no obvious negative features in the difference spectra, but these caveats should generally be borne in mind.

\begin{figure}[htpb]
\plotone{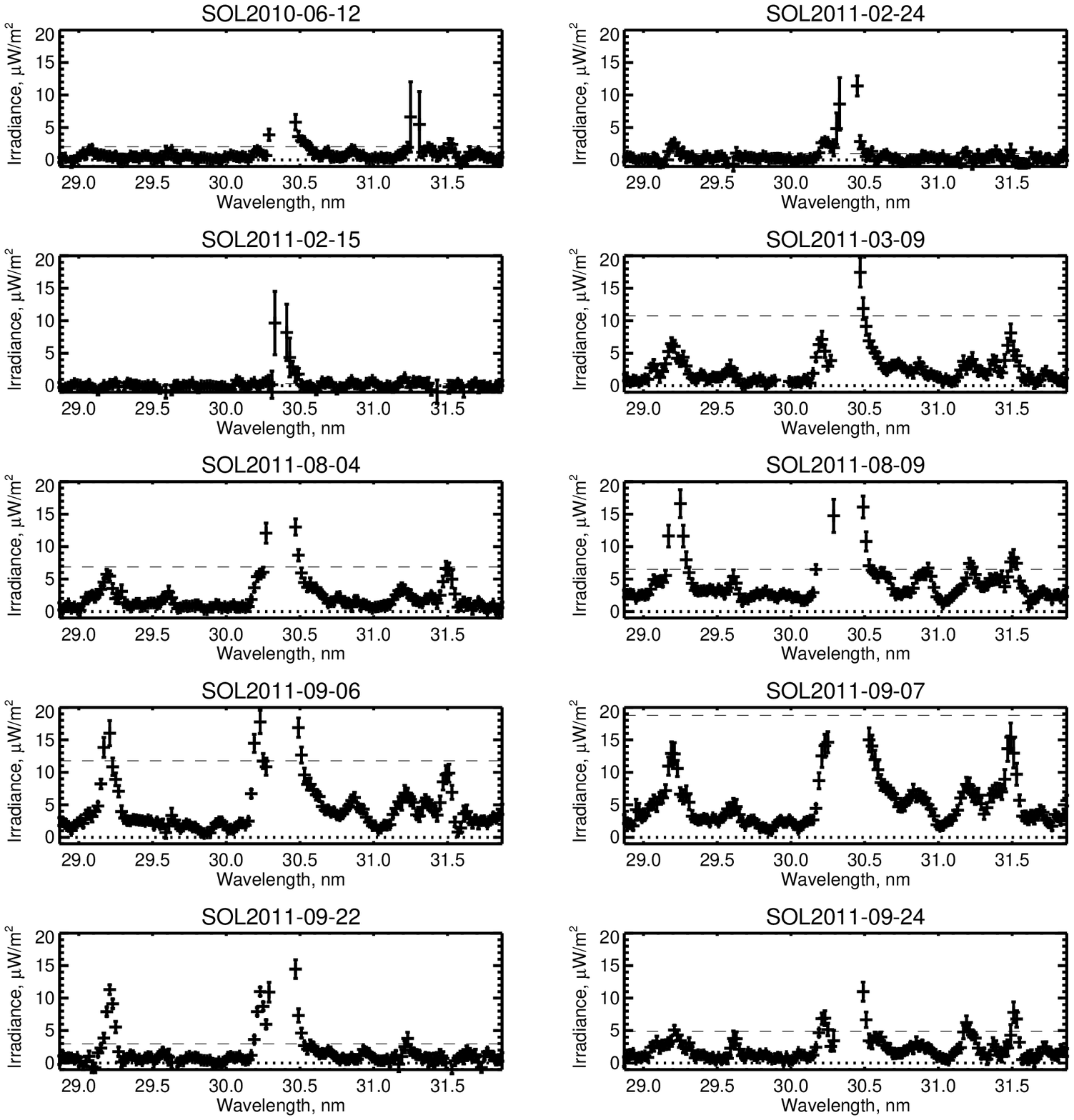}
\caption{Integrated excess irradiance spectra for the impulsive phases of the ten flares, using the start-peak intervals as defined in Table~\ref{tab:times}.
The background intervals are each 110~s beginning at the reference time listed in the table.
The horizontal dotted line shows zero excess, and the dashed lines show 1\% of the peak irradiance in the line
for each integration.
}\label{fig:ten_imp}
\end{figure}

\begin{figure}[htpb]
\plotone{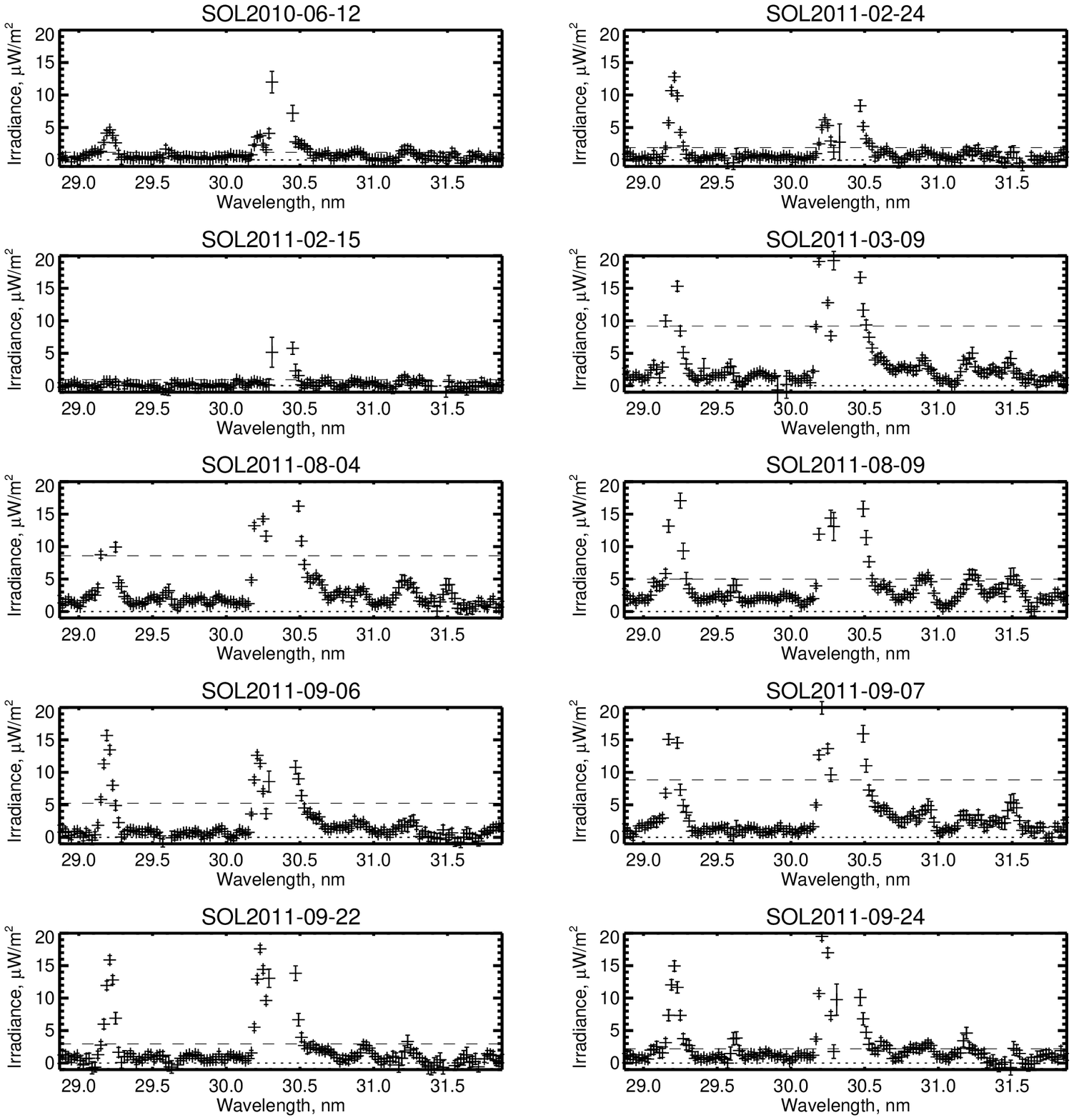}
\caption{Integrated excess irradiance spectra for the gradual phases of the ten flares, using the peak-end intervals as defined in Table~\ref{tab:times}.
The background intervals are each 110~s beginning at the reference time listed in the table.
The horizontal dotted line shows zero excess, and the dashed lines show 1\% of the peak irradiance in the line
for each integration.
}
\label{fig:ten_grad}
\end{figure}

These figures do not show bright  wings of the 30.4~nm line that could be attributed to the expected signature.
In each of the plots the dashed line indicates 1\% of the peak irradiance of the line itself, which is about the level
expected from the calculation of \cite{1990ApJ...351..317P}.
Note that this expected signature would occur mainly in the impulsive phase and (theoretically) should be much less
prominent in the gradual phase.
 In addition the theoretical work assumed a beam intensity much smaller than that necessary to supply the
flare energy, and so the discrepancy is extreme.
All of these events show the presence of non-thermal particles in various ways, including line $\gamma$-radiation and SEPs (but in the case of SOL2011-09-22, only radio signatures), and this set of events therefore provides the best test of this theory thus far possible.
The absence of a charge-exchange signature is somewhat unexpected, noting that \cite{1985ApJ...295..275C} made a
strong case for detectability near Ly$\alpha$.

\section{Discussion}\label{sec:discussion}

Although the charge-exchange signature does not reveal itself here, some of the events do have the appearance of a broad continuum both in the impulsive and gradual phases.
As shown in Figure~\ref{fig:chianti}, the CHIANTI models \citep{1997A&AS..125..149D,2009A&A...498..915D} do show the presence of a broad continuum due to free-free and free-bound contributions, and \cite{2012ApJ...748L..14M} have recently described the EVE continuum spectral contributions for the SOL2011-02-15 event.
\cite{1972SoPh...23..155H} and \cite{1972SoPh...23..169S} showed that the soft X-ray thermal continuum, as observed by GOES, extends into the cm-wavelength range.
Therefore such a component must be present at some level throughout the EUV.
Note that the identification of true continuum regions in the spectrum is not straightforward because of the large numbers of emission lines.
Figure~\ref{fig:chianti_cont} shows \textit{ad hoc} overlays of CHIANTI continua calculated for coronal abundances and
the \cite{1998A&AS..133..403M} ionization-equilibrium tables.
This comparison shows clearly that continuum from hot plasma, at flare temperatures, could not explain the broad-band excess. 
Plasma at lower temperatures (chromospheric/transition region, rather than coronal) could be present, in which case this EUV continuum might eventually reappear in the far infrared/radio domain.
Figure~\ref{fig:chianti} shows that the standard CHIANTI ``flare'' DEM has such a continuum close to the observed levels in our selected continuum bands.

 A pseudo-continuum made up of overlapping emission lines is also a strong possibilty at 1~\AA~resolution.
We note (from Figure~\ref{fig:chianti_cont}) that such a pseudo-continuum may be  prominent below 14~nm, possibly reflecting the density of lines in that  region; the red and blue lines show respectively a GOES-like $10^7$~K thermal continuum, mainly free-free-free emission in the EUV range, and a cooler $10^6$ spectrum for reference.
It is also quite possible that scattered light within the instrument could play a role, but this too would be inconsistent with the appearance of the CHIANTI model continuum in Figure~\ref{fig:chianti}.
A further point requiring investigation would be the possible absorption spectrum of cool intervening material \citep[e.g.,][]{1998SoPh..183..107K}.
Hoever, none of these explanations would affect our conclusions about the charge-exchange radiation, and so lie beyond the scope of the present paper.

\begin{figure}[htpb]
\plotone{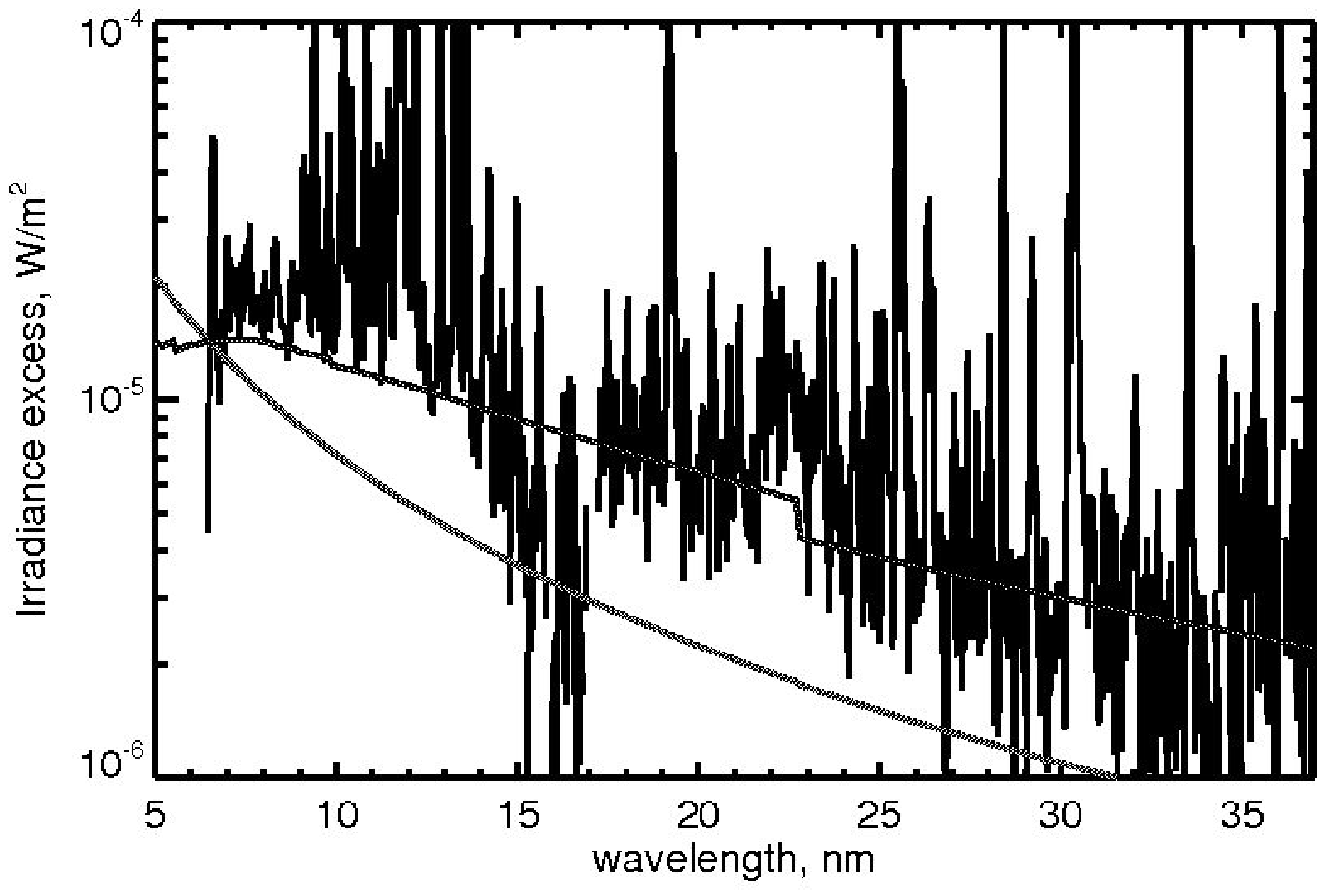}
\caption{EVE/MEGS-A spectrum, 5-37~nm, for the flare excess component of SOL2011-02-15 in the gradual phase (Table~\ref{tab:flares}).
 The lines overlaid are CHIANTI free-free plus free-bound continua for 10$^7$~K (upper) and 10$^6$~K (lower), arbitrarily normalized at the short-wavelength end of the range.
The upper line thus sketches the EUV contribution of the continuum inferred from standard GOES observations.
At the lower temperature the He~{\sc ii} Lyman edge is visible in the model, and perhaps weakly in the data as well.
}\label{fig:chianti_cont}
\end{figure}

We now have the task of understanding why the charge-exchange wings at 30.4~nm do not appear in the EVE spectra.
The flares we have studied definitely have hard X-ray emission and, in some cases, $\gamma$-ray line fluxes.
Generally these flares are the best candidates for the line-wing search.
To function, though, the charge-exchange process requires the penetration of the particle beam into a weakly ionized region.
This requirement conflicts at some level with the fact that flares involve strong heating and ionization of the very atmosphere into which the particles penetrate.
This effect has been considered 
 theoretically and via modeling \citep{1985ApJ...295..275C,1985ApJ...289..425F,1990ApJ...351..317P,1999ApJ...514..430B}, since the original theoretical estimates were based on an equilibrium state within the interacting beam and plasma.
As the beam becomes more intense, this assumption fails, and so there is a time scale for its validity
(see Figure~5 of Canfield \& Chang, and Figure~11 of Peter et al., as well as Table~1 of Brosius \& Woodgate, 1999).
Roughly speaking, the charge-exchange signal will decrease in intensity according to the ratio of the time scale for substantial ionization to be caused by the beam, to the duration of the existence of the beam at a given location.
 For a weak beam structure (a negligible non-Maxwellian tail), as one might expect for example from a spatially diffusive acceleration process, this ratio might remain near unity.
 In the only suggested detection of the Ly$\alpha$ wing signature, the stellar-flare observation of \cite{1992ApJ...397L..95W}, the duration was only a few seconds.

 For a particle interaction highly concentrated into small areas, which is what we observe in hard X-rays \citep[e.g.][]{2006SoPh..234...79H,2011ApJ...739...96K}, this weak-ionization approximation might apply only locally.
The close timing relationship of $\gamma$-ray and hard X-ray emissions in the impulsive phase
\citep[e.g.,][]{1983Natur.305..291F} suggest that proton loss regions (and therefore most likely those of $\alpha$-particles) may be similarly compact.
We presently do not have sufficient angular resolution for $\gamma$-rays  to establish this point directly, but there are strong indications from the \textit{RHESSI} observations that these regions have the double-footpoint property of the hard X-ray sources \citep{2006ApJ...644L..93H}.

The \cite{1990ApJ...351..317P} calculations for the He~{\sc ii} charge-exchange wings assumed a small monoenergetic flux of primary $\alpha$ particles, 5\% by number of a total beam flux of 10$^{13}$~(cm$^2$ s)$^{-1}$, corresponding to about 1\% of the energy of a ``small flare'' of total energy $10^{28}$~erg.
These already result in a peak-to-wing intensity ratio of greater than 1\%, which conflicts directly with our lack of a signature at a much lower level,  and for much more powerful flares.
Further, to interpret the flare $\gamma$-ray observations (including those reflecting primary $\alpha$~particles directly; see e.g. Murphy et al. 1991), a larger $\alpha$ fraction (50\%) is assumed or deduced.
These assumptions together suggest a discrepancy between theory and observation of several orders of magnitude.
We note that the comparison with the peak irradiance of the 304~\AA line is problematic, because even in the
quiet Sun its formation is theoretically challenging \citep[e.g.,][]{2004ApJ...606.1258J}.
But the discrepancy we have found is extreme; the EVE spectra simply show no evidence for the Orrall-Zirker effect.
We cannot describe this more precisely because of the assumptions necessary for the theoretical estimates, which should be tackled again with input from the improved observations.
Among these assumptions we cold also add the primary particle spectral distribution; the beauty of the Orrall-Zirker theory is that we could use charge-exchange reactions to detect low-energy particles.
Indeed, energetic neutral hydrogen atoms from charge-exchange reactions have been detected from solar energetic
protons presumed to interact with He-like ions in the middle corona \citep{2009ApJ...693L..11M}.

\section{Conclusions}

The EVE data make it possible to search for the broad charge-exchange line wings expected in the EUV from a solar flare
that accelerates high-energy $\alpha$-particles, which we expect always to be present in any $\gamma$-ray line event.
We have studied ten flares, including known $\gamma$-ray events and the seven X-class flares from Cycle~24 thus far,
in a search for this process.
There is no charge-exchange signal present in any of these flares, at least according to the signatures predicted by the  existing theoretical work. 
We suggest that further theoretical work may reveal interesting details that explain this discrepancy.
In particular new calculations should deal with the ionization levels more self-consistently.

\bigskip\noindent
{\bf Acknowledgements:} This work was supported by NASA under Contract NAS5-98033 for \textit{RHESSI} for author HSH, who would also like to thank the Physics and Astronomy Group at the University of Glasgow for hospitality.
This work was supported by the Science and Technology Facilities Council grant STFC/F002941/1, by Leverhulme Foundation Grant  F00-179A and by the EU via the HESPE project FP7-2010-SPACE-1/263086.
This work benefited from the hospitality of the International Space Science Institute (Bern), in particular its team activity Flare Chromospheres.
We thank the referee for helping with several clarifications.


\date{\today}

\end{document}